\documentclass[12pt]{iopart}
 \usepackage{graphicx,epsf}
 \newcommand{\be}{\begin{equation}}
 \newcommand{\ee}{\end{equation}}
 \newcommand{\tbar}{\overline{T}}
 \newcommand{\bo}{\mathbf}

\begin{document}
version of May 8, 2012

 \title[Wedges and Cones]
{Wedges, Cones, Cosmic Strings, and the Reality of Vacuum Energy}

 \author
{S. A. Fulling$^{1,2}$, C. S. Trendafilova$^{1,2}$, 
 P. N. Truong$^{1,2.3}$, J.~Wagner$^{1,4}$}

 \address{$^1$ Department of Mathematics,
  Texas A\&M University, College Station, TX, 77843-3368 USA} 

  \address{$^2$ Department of Physics and Astronomy, Texas A\&M University,
 College Station, TX, 77843-4242 USA} 

 \address{$^3$ Present address: Department of Physics, University
 of California, Berkeley, CA, 94720-7300 USA}

 \address{$^4$ Present address: Department of Physics and Astronomy, 
 University of California, Riverside, CA, 92521 USA}

 \ead{fulling@math.tamu.edu}

 \begin{abstract}
One of J.~Stuart Dowker's most significant achievements has been to 
observe that the theory of diffraction by wedges developed a 
century ago by Sommerfeld and others provided the key to solving 
two problems of great interest in general-relativistic quantum 
field theory during the last quarter of the twentieth century:  the 
vacuum energy associated with an infinitely thin, straight cosmic 
string, and (after an interchange of time with a space coordinate) 
 the apparent vacuum energy of empty space as viewed by 
an accelerating observer.  
 In a sense the string problem is more elementary than the wedge,
  since 
Sommerfeld's technique was to relate the wedge problem to that of a 
conical manifold by the method of images.
 Indeed, Minkowski space, as well as all cone and wedge problems, 
are related by images to an infinitely sheeted master manifold, 
which we call Dowker space.
  We review the research in this area
 and exhibit in detail the vacuum expectation values of the energy 
density and pressure of a scalar field in Dowker space and the cone 
and wedge spaces that result from it.
 We point out that the (vanishing) vacuum energy of Minkowski 
space results, from the point of view of Dowker space, from the
quantization of angular modes, in precisely the way that the 
Casimir energy of a toroidal closed universe results from the 
quantization of Fourier modes; we hope that this understanding 
dispels any lingering  doubts about the reality of cosmological 
vacuum energy.
\end{abstract}

 \pacs{03.70.+k, 41.20.Cv}
  \ams{81T55, 34B27, 81Q05}
 \maketitle

\section{Introduction}\label{wcintro}

 \subsection{Basics} 
\label{basics}

 A \emph{wedge} is a region  
 $\Omega \subset \bo R^3$ bounded by two intersecting 
planes on which, say, the Dirichlet boundary condition is imposed.  
All the standard linear partial differential equations (wave, heat, 
Schr\"odinger, resolvent, \dots) for all the standard fields (scalar, 
electromagnetic, spinor, \dots) can be studied there, and their 
respective Green functions constructed, by standard but not totally 
trivial methods.

 As the prototype of a wedge problem we consider  the conditions 
defining the ``cylinder kernel''\negthinspace,
  which provides  \cite{lukosz2,L,BH,systemat} the 
most direct way of calculating the vacuum expectation value of the 
energy  of a quantized massless scalar field:
  \be  T(t,\mathbf{x},\mathbf{x}') \quad
 \hbox{is defined in $\mathbf{R}^+\times\Omega\times\Omega$,} 
 \label{Tdomain}\ee
 \be {\partial {^2T}\over\partial{t^2}} = 
 - \nabla^2 T \quad\hbox{for $\mathbf{x}$ in $\Omega$,} 
 \label{TPDE}\ee 
 \be  T(t,\mathbf{x},\mathbf{x}')=0 \quad
 \hbox{for $\mathbf{x}$ in $\partial\Omega$,}\label{TBC} \ee 
\be  T(0,\mathbf{x},\mathbf{x}')=\delta(\mathbf{x} -\mathbf{x}') 
   \quad\hbox{for $\mathbf{x}$ in $\Omega$}, \label{TIC}\ee 
 \be  T(t,\mathbf{x},\mathbf{x}')\quad\hbox{is bounded as $t\to+\infty$}. 
 \label{Tdecay}\ee 
Explicitly, in the three-dimensional case, we have
 \be  \Omega = \{(r,\theta,z)\colon  0<r<\infty, \> 
0<\theta<\theta_0\,, \> -\infty<z<\infty \}, \label{3domain}\ee 
\be  \nabla^2= {\partial{^2}\over\partial{r^2}}
  +\frac1r \,{\partial{}\over\partial{r}} +\frac1{r^2}\, 
{\partial{}\over\partial{\theta^2}} +{\partial{^2}\over\partial{z^2}}
 \,,\label{3Lap}\ee 
\be \delta(\mathbf{x} -\mathbf{x}') 
 = \frac1{r}\,\delta(r-r')\delta(\theta-\theta')\delta(z-z') ,
\label{3delta}\ee 
 and an additional implied boundary condition,
 \be  T(t,\mathbf{x},\mathbf{x}')\quad\hbox{is bounded as 
$z\to\pm\infty$}. \label{3decay} \ee 
Often the 
two-dimensional reduction is of interest:
 \be  \Omega = \{(r,\theta)\colon  0<r<\infty, \> 
0<\theta<\theta_0 \}, \label{2domain} \ee 
\be  \nabla^2= {\partial{^2}\over\partial{r^2}}
  +\frac1r \,{\partial{}\over\partial{r}} +\frac1{r^2}\, 
{\partial{}\over\partial{\theta^2}}\,, \label{2Lap}\ee 
 \be \delta(\mathbf{x} -\mathbf{x}') 
 =  \frac1r\,\delta(r-r')\delta(\theta-\theta').\label{2delta} \ee

 \emph{Remarks:}  The corresponding Green function for all of $\bo 
R^3$ in the role of~$\Omega$ is 
 \be
 T_0(t,\bo x,\bo x') = 
 \frac{t}{\pi^2(t^2+ |\bo x-\bo x'|^2)^2}\,.
 \label{T0}\ee
 It is proportional to the $t$-derivative of the 
 fundamental solution of the Laplacian 
in~$\bo R^4$, with $t'=0$.
 It follows that the most relevant dimension is~$4$, the same as 
the space-time dimension of the quantum field theory, and we 
henceforth revise our terminology accordingly.
 It is important to understand that in the cylinder-kernel 
formalism, $t$ is ``imaginary'' time, related to the physical time 
$x^0$ by $it= x^0-x^{0\prime}$.
 The term ``cylinder kernel'' refers to the domain $\mathbf R^+ 
\times \Omega$ in $\mathbf R^4$, not to the cylindrical character 
of $\Omega$ itself in the present work.

 The angle $\theta_0$ is ``good'' if $\theta_0=\pi/N$ for some 
integer~$N$.
In other words, the wedge considered is one of $2N$ equal sectors into 
which Euclidean space is divided.
 (For mixed boundary conditions 
 (Neumann on one side and Dirichlet on the other), 
 $\theta_0$ is good only if $N$ is 
even (the number of sectors is a multiple of~$4$).)
In such a case the problem can be solved immediately by the method 
of images:  Replicate the singular source \eref{3delta} by 
perpendicular reflection of $\mathbf{x}'$
 through  each boundary plane, 
 with a minus sign for the Dirichlet condition but a plus for 
Neumann.  Continue the construction through all $2N$ sectors;
 the two directions of continuation meet consistently.
 The solution of the PDE in $\bo R^3$ with this multiple source 
 is easily constructed as a linear combination of copies of $T_0$ 
 with its source moved to each of the images.
That function satisfies the desired boundary condition on each plane, so
its  restriction to~$\Omega$ is the desired Green function~$T$.

If the angle is ``bad''\negthinspace, the replicated wedges cannot 
match up to constitute Euclidean space.
 The constructed covering space is a \emph{cone}, which is locally 
flat but has a singularity at the axis.  The circumference of a 
circle centered at the axis is $\theta_1$ times the radius, where 
$\theta_1$ is $\theta_0$ times the number of sectors in the 
construction.
 (The number of sectors in such a case is not uniquely determined.
 One could either take the simplest case,  two sectors, or choose the 
value that makes $\theta_1$ closest to~$2\pi$.) 
 To solve  a wedge with a bad angle by images, therefore, one 
must know the cylinder kernel for the related problem on the cone.
The latter satisfies the same equations \eref{TPDE}, \eref{TIC},
 \eref{Tdecay}, \eref{3decay}
 ($\Omega$ now being the cone manifold), but the boundary 
condition \eref{TBC} is replaced by the periodicity condition
 \be\fl
  T(t,r,\theta+\theta_1,z,\mathbf{x}')= T(t,r,\theta,z,\mathbf{x}'),
 \qquad
{\partial {T}\over\partial{\theta}} (t,r,\theta+\theta_1,z,\mathbf{x}')= 
{ \partial {T}\over\partial{\theta}}(t,r,\theta,z,\mathbf{x}') .
 \label{3period}\ee
 Once this Green function has been found, it can be used in the 
role of~$T_0$ to construct the $T$ for the wedge.

 In both cone and wedge, there is a technical issue about 
uniqueness of the solution, which is resolved by choosing the 
solution of minimal growth as the axis is approached~\cite{meix,KS}.

 \subsection{History} \label{history}

At the end of the 19th century Sommerfeld 
\cite{som96,som97,som01,som35} used the method of images to 
solve the problem of diffraction of waves by edges and wedges.
 (The seminal paper \cite{som96} has been translated with extensive 
commentary in~\cite{som04}.)
 His and most other early work was concerned with the wave equation 
and with the solution corresponding to a particular incident wave, 
but of course the foregoing remarks about boundary conditions and 
images also apply there.
 At that time it was natural to think of the cone as a Riemann 
surface in the sense of complex analysis, rather than a locally 
flat Riemannian manifold in the sense of differential geometry.
 Therefore, initially only values of $\theta_1$ of the form 
 $2\pi M$ were considered, and hence wedges of angles
 $\theta_0=\pi M/N$.
 Sommerfeld \cite{som96} was primarily concerned with a
  sharp edge (conducting half-plane), for which $\theta_0=2\pi$ 
and hence a $2$-sheeted Riemann surface ($\theta_1=4\pi$)
divided into two sectors ($N=1$) suffice.
 He gave mere indications of how to treat  the general case 
 $\theta_0=\pi M/N$, but his method was made explicit by 
Carslaw~\cite{car99}, and the case of a right-angle wedge 
($\theta_0={3\pi\over 2}$, three sheets, four sectors)
  was worked out in detail by Reiche~\cite{reiche}. 
   (The case $2\pi-\theta_0={\pi\over 3}$, 
$M=5$, $2N=6$, is treated in the notes to~\cite{som04}.)
Sommerfeld \cite[pp 346--347, 357]{som96} also showed how to 
construct solutions on an infinite-sheeted Riemann surface, but did
not really explain how to use them to solve the original problem for 
 $\theta_0\ne \pi M/N$ (and  the commentators in~\cite{som04} 
describe this passage as ``surrealistic'').
 He gave an explicit, though still very brief, prescription in 
\cite[p 38]{som01}, as cited in Carslaw~\cite{car20}, and Wiegrefe 
\cite{wieg} gave a more thorough treatment.

 Sommerfeld constructed solutions on Riemann surfaces by conformal 
mappings from the complex plane, but then he represented them by 
certain contour integrals,
 which are periodic with period~$\theta_1\,$.  
 Carslaw \cite{car10,car20} noted that these 
integrals continue to satisfy 
the desired differential equations and periodicity
  even when $\theta_1/\pi$ is not rational. 
  Knowing this, it is possible 
to dispense with the Riemann surfaces and obtain the solution of 
the wedge problem 
 --- for arbitrary $\theta_0= \frac12 \theta_1$  ---  
 from this periodic function by adding a single image term.
 Carslaw's papers \cite{car99,car10,car20} are also notable for 
 emphasizing the generality of the method, beyond the wave 
equation.
 Other work on wedge  problems during this era is 
reviewed in the introduction to \cite{O}, which also constructs
Green functions for the wave equation for 
arbitrary~$\theta_0$ in terms of standard special functions, 
without relying on either Riemann surfaces or contour integral 
representations. 
 See also \cite{J1,J2,SPS} for applications of concepts of 
scattering and diffraction theory.

Until 1977 the mathematically simpler problem of a cone was thought 
of by physicists merely as a means to the end of solving the 
problem of a wedge;
 it would never have occurred to  Sommerfeld to speak of a ``cosmic 
string''\negthinspace, for example.
 But Stuart Dowker was well versed in this classic literature,
 and he realized that it could be instantly applied to problems 
that had arisen in quantum field theory in curved space-time 
\cite{D77,D78,D87,D87a,D90}.
  Again, the simplest problem was originally seen as a tool for 
  studying something more recondite:
 In \cite{D77,D78} the emphasis is on the analytic continuation  of 
a cone manifold to ``Rindler space''\negthinspace, the space-time 
visible to a uniformly accelerated observer; the periodicity is in 
the (imaginary) time coordinate and represents the effective 
temperature of the vacuum state.  (In lieu of a long list of 
references, we refer to a previous review article~\cite{FR}.)
 Ten years later, however, the cone as a spatial manifold came to 
be regarded as physically realistic, as the idealized zero-radius 
limit of a cosmic string. (For subtleties in this interpretation 
see \cite{GT,FG}.) 
 The Green functions and the vacuum expectation values of the stress 
tensor near a straight cosmic string were calculated in 
\cite{D87,D87a} and many other papers, including 
 \cite{HK,smith,FS,FPZ,parker,DS,SH,L87,L92,L95,L96,GL}.

In addition to  moving the focus from wedges to cones, 
Dowker \cite{D77} introduced a  conceptual shift:
 The infinite-sheeted Riemann surface~${\cal M}_\infty$,
  previously regarded as a 
necessary evil required only for irrational cone angles, is now 
regarded as the fundamental manifold from which the others are 
built. 
 (See also \cite{Zio} and references therein.)
  A $\theta_1$-periodic image sum 
 (of periodically displaced copies  of 
a Green function on~${\cal M}_\infty$)
yields the corresponding Green function on ${\cal M}_{\theta_1}\,$,
 the cone of defining angle~$\theta_1\,$. 
 This construction (which we discuss in more detail in 
\sref{periodic})
 is precisely analogous to the creation of Green 
functions on an interval (or in a rectangular box) as periodic 
image sums of Green functions on the real line 
 (or a higher-dimensional Euclidean space).
 A Green function for a wedge is then obtained by one more step of 
image summation, this time involving a single reflected copy
 (or, more generally, a sum over some even number of images,
including the original  source point).
 In the special case $\theta_1=2\pi$, of course, the Green function 
for Minkowski space is recovered.
 Although the Green function on      ${\cal M}_\infty$ is harder to 
calculate than that on ${\cal M}_{2\pi}$ and perhaps not 
particularly easier than that  on a general ${\cal M}_{\theta_1}\,$,
 the picture in which the covering space ${\cal M}_\infty$ is 
 the basic object is nicely consistent with contemporary notions of 
path integrals (as formal representations of exact solutions) and 
classical-path sums (as the basic framework for high-frequency and 
semiclassical approximations).
It seems appropriate to refer to ${\cal M}_\infty$ 
 (including any extra spatial coordinates such as $z$ 
 in~\eref{3domain})
 as the \emph{Dowker manifold} and to the corresponding space-time 
model as \emph{Dowker space}.

 Dowker worked primarily with the same contour-integral 
representations as Sommerfeld and Carslaw.
Our present exposition is strongly influenced by a
  later paper by Smith \cite{smith},  which derives the cylinder 
  kernels in   total dimensions 3 and 4 
 without overt use of such 
  complex analysis (``overt'' because Smith, and we, freely use 
  special-function identities that appear in standard handbooks 
  without inquiring how they were proved). 
 Formulas in different dimensions are related in a way 
systematically investigated by Guimar\~aes and Linet \cite{GL},
 which we shall explain in due course.
Another method leading to the same results for the cone 
 is that of Helliwell and Konkowski \cite{HK}, 
analogous to
earlier work for the wedge included in the famous 
paper~\cite{DeuC}.

 Finally, we must step back to an earlier paper by Lukosz~\cite{L},
which has been largely overlooked by authors in relativity until 
recently.  This paper deals with the cylinder kernel (and vacuum 
energy) for a wedge.  It is based on an insight similar to 
Carslaw's \cite{car10,car20}.  Lukosz uses the method of images to 
solve the problem for the very special cases $\theta_0=\pi/N$, as 
in \sref{basics}. He works the solution into a closed form (i.e., 
one no longer involving a sum over sectors) and then observes that 
the result satisfies the conditions of the problem for \emph{all} 
values of $\theta_0\,$.  His method of course gives also the 
solution for an arbitrary cone.  It can be checked to agree with 
those of \cite{D78,L87,smith}, all somewhat differently derived.

 Papers continue to be written about vacuum energy in the presence 
of either wedges or cosmic strings, the scenario being complicated 
or generalized in various ways of greater or lesser physical 
importance.  An undoubtedly incomplete list is
\cite{parkerstr,bordag2str,G,KB,RS,ST,BMBST,BMBS,BS,OT,SV,MWK,dowsphpis}.
 The most experimentally relevant generalization is to the 
electromagnetic field, whose study goes back to Deutsch and 
Candelas \cite{DeuC} and has been continued in 
 \cite{BL,NLS,BEM,EBM} and numerous other references that can be 
traced back from \cite{BEM,EBM}.

\section{Formalism in cylindrical coordinates} \label{cylform}

In order to calculate the vacuum expectation values of all  
 the components of the stress-energy tensor, $T_{\mu\nu}\,$, 
 we use a different 
 cylinder kernel, $\tbar$, which is an antiderivative of $T$ as 
defined in \sref{basics}.  $\tbar$ satisfies equations 
 \eref{Tdomain}--\eref{3decay} except that \eref{TIC} is replaced by
 \be
\frac{\partial\tbar}{\partial t}(0,\bo x,\bo x')=
 \delta(\bo x-\bo x').
\label{TbarIC} \ee
  In 
the calculations that follow, either ``$0$" or an overdot
 is used to refer to components 
and derivatives with respect to the physical time coordinate.

The  stress-energy tensor formula 
 for a massless real scalar field in flat space is  \cite{spo}
\begin{eqnarray}
 T_{\mu\nu} &= \partial_\mu \phi \partial_\nu \phi - 
\frac{1}{2}\eta_{\mu\nu} \partial^\lambda \phi \partial_\lambda 
\phi + \xi[\eta_{\mu\nu} \partial_\lambda \partial^\lambda(\phi^2)-
\partial_\mu \partial_\nu (\phi ^2)] \nonumber \\
 &= \frac12 \partial_\mu \phi \partial_\nu \phi
  - \frac12\phi\partial_\mu\partial_\nu\phi
+ \beta[\eta_{\mu\nu} \partial_\lambda \partial^\lambda(\phi^2)-
\partial_\mu \partial_\nu (\phi ^2)]
 \label{st}\end{eqnarray}
 in Cartesian coordinates.
       Here $\xi$ is the usual curvature coupling parameter, but
 we write  $\xi = \beta + \frac{1}{4}$ because then $\beta=0$  
 is the algebraically simplest case;
 thus $\beta=-\frac14$ for minimal coupling and $\beta=-\frac1{12}$ 
for conformal coupling in dimension $3+1$.
 In the second version of \eref{st} we have dropped a term that 
vanishes by virtue of the equation of motion, 
$\partial_\lambda\partial^\lambda\phi=0$.

To calculate the components of the tensor in  
 cylindrical coordinates, one can use \eref{st} 
 with $\eta$ and $\partial$
 reinterpreted as the metric tensor and covariant derivative.
 Alternatively, one can proceed as
 Schwartz-Perlov and Olum did for  spherical 
coordinates~\cite{spo}:
 Without loss of generality, calculate at a point where 
 the $r$, $\theta$, and $z$ unit 
vectors point along the $x$, $y$, and $z$ axes.
 Define the  ``$\perp$" components of tensors to 
be along the $\theta$ direction, but with respect to an orthonormal basis,
 so that they  have the same physical units as the other 
components; thus on any scalar function
 \be
 \partial_\bot  \phi =\frac 1r \, \partial_\theta \phi,
\label{thd1} \ee
and in \eref{st} the metric is $\eta_{00}=-1$, 
$\eta_{rr}=\eta_{\perp\perp}=\eta_{zz}=1$, all other 
components~$0$. 
 The only remaining complication is that
  in the terms  involving second-order derivatives the derivatives 
of the local basis vectors must be taken into account.
 That can be done by means of Christoffel symbols as usual in 
relativity or by 
direct calculation as in elementary vector calculus, with the 
results
 \begin{equation}
\partial_\perp{}\!^2\phi=\frac{1}{r}\,\partial_r \phi+
 \frac{1}{r^2}\,\partial_\theta{}\!^2\phi
 \label{perpderivs}\end{equation} 
and \eref{mixedderiv}.
Note also that
 \be
 \partial_\theta{}\!^2(\phi^2) = 2(\partial_\theta\phi)^2
 +2\phi\partial_\theta{}\!^2\phi,
\label{phisq} \ee
 so that any one of $\partial_\theta{}\!^2(\phi^2)$, 
 $(\partial _\theta\phi)^2$, and $\phi \partial_\theta{}\!^2\phi$
 can be eliminated in favor of the other two.
 Continuing to follow \cite{spo}, one then notes that,
 because of the symmetry of the situation, the expectation value 
of $\phi^2$ does not depend on time or 
 (in the absence of end plates) the $z$ coordinate, so 
$\partial_0{}\!^2\langle\phi^2\rangle=\partial_z{}\!
^2\langle\phi^2\rangle=0$. 
 For Dowker space or a cone (but not for a wedge)
 the expectation value of $\phi^2$ also does not depend 
on~$\theta$.
 All these simplifications yield
 \begin{eqnarray}
 T_{00}=\frac{1}{2}[\dot{\phi}^2-\phi\ddot{\phi}]-
2\beta[\partial_r(\phi\partial_r\phi)+\frac{1}{r}\phi\partial_r\phi] 
,
\label{sol31}\\ 
 T_{rr}=\frac{1}{2}[(\partial_r\phi)^2-
\phi\partial_r ^2\phi]+2\beta[\frac{1}{r}\phi\partial_r \phi] ,
 \label{sol32}\\ 
T_{\perp\perp}=
 \frac1{2r^2} Q(\phi)
 -\frac{1}{2r}\phi\partial_r\phi
  +2\beta[\partial_r(\phi\partial_r \phi)], \label{sol33}\\ 
T_{zz}=\frac{1}{2}[(\partial_z\phi)^2-\phi\partial_z 
^2\phi]+2\beta[\partial_r(\phi\partial_r\phi)+\frac{1}{r}\phi\partial_r\phi] 
,\label{sol34}
 \end{eqnarray}
 where
 \be
Q(\phi) = (\partial_\theta\phi)^2 -\phi \partial_\theta{}\!^2 \phi
 =2(\partial_\theta\phi)^2 - \frac12\partial_\theta{}\!^2(\phi^2).
 \label{perpalts}\ee

 Now we can rewrite the components of $T_{\mu\nu}$ in terms of the 
cylinder kernel, $\tbar$, rather than $\phi$. Since $\langle 
0|\phi(x)\phi(x')|0\rangle=-\frac{1}{2}\tbar(x,x')$, we have,
 for example,  
\be \textstyle
 \langle\phi\partial_r\phi\rangle=-\frac{1}{2}\partial_r\tbar, 
\quad
\langle\phi\partial_r ^2\phi\rangle=-\frac{1}{2}\partial_r 
^2\tbar, \quad
\langle(\partial_r\phi)^2\rangle=-
\frac{1}{2}\partial_{r'}\partial_{r}\tbar,
 \ee
where $x'$ is set equal to $x$ after differentiation.
 (Strictly speaking, 
 we should symmetrize $2\partial_r{}\!^2$ as
 $\partial_r{}\!^2 + \partial_{r'}\!^2$, etc., but it can be seen 
that this does not matter.)
 The derivatives for the other coordinates are computed similarly. 
 The components of $T_{\mu\nu}$ finally become
\begin{eqnarray}
\langle T_{00}\rangle&=-\frac{1}{2}\partial_t   
^2\tbar+\beta[\partial_r\partial_{r'}\tbar+\partial_r 
^2\tbar+\frac{1}{r}\partial_r\tbar] \label{sol41}\\
\langle  T_{rr}\rangle&=-
\frac{1}{4}[\partial_r\partial_{r'}\tbar-\partial_r ^2\tbar]-
\frac{1}{r}\beta\partial_r\tbar, \label{sol42} \\ 
\langle T_{\perp\perp}\rangle&=\frac{1}{4r}\partial_r\tbar
+\frac1{4r^2}[\partial_\theta{}\!^2\tbar-\partial_\theta 
\partial_{\theta'}\tbar]
-\beta[\partial_r\partial_{r'}\tbar+\partial_r\tbar]
 \quad\hbox{(general)}\label{sol43a}\\  
 &=\frac{1}{4r}\partial_r\tbar+\frac{1}{2r^2}\partial_\theta 
^2\tbar-\beta[\partial_r\partial_{r'}\tbar+\partial_r\tbar]
 \quad\hbox{(symmetric case)},\label{sol43b}\\ 
\langle T_{zz}\rangle&
 =-\frac{1}{4}[\partial_z\partial_{z'}\tbar-\partial_z 
^2\tbar]-\beta[\partial_r\partial_{r'}\tbar+\partial_r 
^2\tbar+\frac{1}{r}\partial_r\tbar] \label{sol44}. 
 \end{eqnarray}
 Henceforth the expectation-value brackets around components of the 
stress tensor will be omitted for simplicity of notation; and 
later the notation $T_{\mu\nu}$ will be abused once again 
after a vacuum subtraction.
 Again in these equations it is understood that formally the two 
space-time points are set equal after the differentiations.
 In reality these diagonal values are divergent, and either the 
points must be kept separated until after a vacuum subtraction has 
been performed, or some other kind of regularization must be 
adopted.

 In the interior of a wedge it is possible for $T_{r\bot}$ to have 
a nonzero vacuum expectation value, but we shall not consider that 
component of the tensor further in this paper except to record 
the needed formula
\be
 \partial_\bot\partial_\|\phi = \partial_\|\partial_\bot\phi
 = \frac1r\, \partial_r\partial_\theta \phi 
  - \frac1{r^2}\,\partial_\theta \phi.
 \label{mixedderiv}
\ee
Here, to prevent ambiguity we write $\partial_\|$ for the 
radial covariant derivative.  (When it stands alone, $\partial_\|$
is equivalent to $\partial_r\,$:  $\partial_\|{}\!^2\phi = 
\partial_r{}\!^2\phi$, for instance.)

\section{Calculation of $\tbar$} \label{tbarcalc}

We now consider the problem of calculating  cylinder kernels
  in polar coordinates. 
 Various methods can be used.
 Guided by physical motivations, Smith \cite{smith} used one method 
in three (total) dimensions and a different one in four.
 We have carried out both calculations in both ways, and here we 
demonstrate how to do the four-dimensional case each way, to stress 
that they are mathematically interchangeable.
 Finally, we summarize the three-dimensional results and describe 
how adjacent dimensions are related.

 \subsection{Homogeneous boundary-value problem; $r$--$\theta$--$z$ 
expansion}
The first method is to solve the boundary-value problem set up in 
 \sref{wcintro} (or its modification by~\eref{TbarIC}).
Thus $\tbar$ satisfies for $t>0$ 
  the homogeneous 4-dimensional Laplace equation,
  \begin{equation}
\label{eq:drv}\frac{\partial ^{2}T}{\partial t^{2}} + 
\frac{\partial ^{2}T}{\partial r^{2}} + \frac{1}{r}\frac{\partial 
T}{\partial r} +\frac{1}{r^{2}} \frac{\partial ^{2}T}{\partial 
\theta ^{2}} + \frac{\partial^{2}T}{\partial z^{2}}= 0,
\end{equation} 
 along with a periodicity condition,
  $T(\theta + \theta_{1}) = T(\theta)$, and
 the intial condition
\begin{equation}
\label{eq:T0}
T(0, \textbf{r}, \textbf{r}') = \frac{1}{r}\delta(r-r')
 \delta(\theta -\theta')\delta(z-z').
\end{equation} 
 Expanding $T$ in a Fourier series in $\theta$ yields
\begin{equation}
 T(t,r,\theta,z) = \sum^{\infty}_{n = -\infty} 
e^{in\theta(\frac{2\pi}{\theta_{1}})} T_{n}(t,r,z) \end{equation} 
with 
 \begin{eqnarray}
\label{eq:intT}
 T_{n}(t,r,z) &=& \frac{1}{\theta_{1}} \int^{\theta_{1}}_{0} e^{-
in\theta(\frac{2\pi}{\theta_{1}})} T(t,r,\theta,z)\,d\theta. 
\end{eqnarray}
  Using these expressions in (\ref{eq:drv}), we see that
\begin{equation}
\label{eq:deriv}
\frac{\partial ^{2}T_{n}}{\partial t^{2}} + \frac{\partial ^{2}T_{n}}
 {\partial r^{2}} + \frac{1}{r}\frac{\partial T_{n}}{\partial r}
  -\frac{n^{2}}{r^{2}}\left(\frac{2\pi}{\theta_{1}}\right)^{2}T_{n} + 
\frac{\partial ^{2}T_{n}}{\partial z^{2}}= 0.
\end{equation}
  From (\ref{eq:T0}) and (\ref{eq:intT})
 we obtain
\begin{eqnarray}
T_{n}(0,r,z) &=& \frac{1}{\theta_{1}}\int^{\theta_{1}}_{0} d\theta 
 e^{-in\theta(\frac{2\pi}{\theta_{1}})}
  \frac{1}{r}\delta(r-r')\delta(\theta -\theta')\delta(z-z')
  \nonumber\\
\label{eq:Tn0a}
&=& \frac{1}{\theta_{1}}e^{-in\theta'(\frac{2\pi}{\theta_{1}})}
 \frac{1}{r}\delta(r-r')\delta(z-z').
\end{eqnarray}

We now make a further step of variable separation,
  $T_{\rm{sep}}(t,r) = T(t)R(r)Z(z)$, and we define $\lambda 
= \frac{2n\pi}{\theta_{1}}$
 and occasionally suppress ``$n$'' in the notation.
  From (\ref{eq:deriv}) we get 
 \begin{eqnarray}\frac{T''}{T} + \frac{R''}{R} + \frac{Z''}{Z} +
  \frac{1}{r}\frac{R'}{R} - \frac{\lambda^{2}}{r^{2}} = 0.
\end{eqnarray} We let 
\begin{eqnarray}
-\frac{T''}{T} - \frac{Z''}{Z} &=& \frac{R''}{R} + \frac{1}{r}\frac{R'}{R}
  - \frac{\lambda^{2}}{r^{2}} 
 = -\omega^2,
\end{eqnarray} 
 and therefore conclude that
\begin{equation}
\label{eq:Bessel}
 R'' + \frac{1}{r}R' + \left(\omega^2
  - \frac{\lambda^{2}}{r^{2}}\right)R = 0, 
 \ee
 \be Z = e^{ikz}, \qquad
 T = e^{-\omega' t}, \quad
 \omega^{\prime2} \equiv \omega^2 + k^2.
  \label{eq:ZT}
 \ee
The appropriate solution of (\ref{eq:Bessel}) is the Bessel function
 $J_{|\lambda|}(\omega r)$ (and not $Y_{|\lambda|}(\omega r)$ 
 because of the minimal irregularity at $r=0$). 
 Now we can define the combined Bessel and Fourier transform
\begin{equation}
\label{eq:Tn0b}
T_{n}(t,r,z) = \int^{\infty}_{0}\omega \,d\omega \int^\infty_{-\infty}
  dk\, \tilde{T}(\omega,k) J_{|\lambda|}(\omega r)e^{-\omega' t}e^{ikz}.
\end{equation} When $t=0$, from (\ref{eq:Tn0a}) we get
\begin{equation}
T_{n}(0,r,z) = \frac{e^{-i\lambda\theta'}\delta(r-r')\delta(z-z')}
 {\theta_{1}r} \equiv  P(r,z),
\end{equation}
  and from (\ref{eq:Tn0b}) we see that
\begin{equation}
T_{n}(0,r,z) = \int^{\infty}_{0}\omega\, d\omega \int^\infty_{-
\infty} dk\, 
 \tilde{T}(\omega,k) J_{|\lambda|}(\omega r)e^{ikz}.
\end{equation}
  Therefore we can solve for $\tilde{T}(\omega,k)$:
\begin{eqnarray}
\tilde{T}(\omega,k) &=& \frac{1}{2\pi}\int^\infty_{-\infty}dz 
 \int^{\infty}_{0}r\,dr\,J_{|\lambda|}(\omega r)P(r,z)e^{-ikz}
  \nonumber\\
&=& \frac{e^{-i\lambda\theta'}}{2\pi\theta_{1}}J_{|\lambda|}
 (\omega r')e^{-ikz'}.
\label{Ttilde}\end{eqnarray}
  From this one finally gets
\begin{equation}
T(t,r,\theta,z) = \int^{\infty}_{0}\omega\, d\omega 
 \int^\infty_{-\infty} dk \sum^{\infty}_{n=-\infty} 
 \tilde{T}_{n}(\omega,k) J_{|\lambda|}(\omega r)
 e^{in\theta}e^{-\omega' t}e^{ikz} 
\label{noglory} \ee
 or, in full glory,
  \begin{eqnarray}
T(t,r,\theta,z,r',\theta',z') &=& \frac{1}{2\pi\theta_1}
  \sum_{n=-\infty}^\infty \int_{0}^{\infty} \omega\, d\omega 
 \int^\infty_{-\infty} dk 
J_{|\lambda|}(\omega r) J_{|\lambda|}(\omega r') \nonumber\\
&& \times e^{i\lambda(\theta-\theta')} e^{-\omega' t} e^{ik(z-z')}.
\label{glory}\end{eqnarray} 

 The formula for $\tbar$ is the same as for $T$ except for a 
factor $-\omega'$ in the denominator to implement the integration 
with respect to~$t$:
\begin{eqnarray}
\tbar(t,r,\theta,z,r',\theta',z') &=&-\, \frac{1}{\pi\theta_1}
  \sum_{n=-\infty}^\infty e^{i\lambda(\theta-\theta')}
  \int_{0}^{\infty} \omega\, d\omega\, J_{|\lambda|}(\omega r) 
 J_{|\lambda|}(\omega r') \nonumber \\
&& \times  \int^\infty_{-\infty}\, dk \, (\omega^2+k^2)^{-1/2}
 e^{-t\sqrt{\omega^2 + k^2} } e^{ik(z-z')}.
\label{tbarinitial}\end{eqnarray} 
By \cite[(3.961.2)]{GR}  the $k$ integral 
 equals $2K_0(\omega \zeta)$, where 
 $\zeta=\sqrt{t^2 + (z-z')^2}$. Hence we have 
\begin{equation}\fl
\tbar(t,r,\theta,z,r',\theta',z') = -\,\frac{1}{\pi\theta_1}
  \sum_{n=-\infty}^\infty e^{i\lambda(\theta-\theta')}
  \int_{0}^{\infty} \omega\, d\omega\, J_{|\lambda|}(\omega r) 
 J_{|\lambda|}(\omega r')K_0(\omega \zeta).
\end{equation}
  The integral over $\omega$ can be computed
 \cite[(6.522.3)]{GR}, giving us the 
 Fourier-series representation of $\tbar$,
\begin{eqnarray}
\tbar(t,r,\theta,z,r',\theta',z') &=& -\frac{1}{\pi\theta_1} 
 \sum_{n=-\infty}^\infty e^{i\lambda(\theta-\theta')} 
 \frac{1}{r_1 r_2}\left(\frac{r_2 - r_1}{r_2 + r_1}\right) ^{|\lambda|},
\end{eqnarray} where $r_1 \equiv \sqrt{(r-r')^2 + \zeta^2}$
  and $r_2 \equiv \sqrt{(r+r')^2 + \zeta^2}$. 
 Let 
 \be \frac{r_2 - r_1}{r_2 + r_1} \equiv e^{-u}. \label{udef1} \ee
  Smith \cite{smith} points out the alternative formulas for $u$ 
  that we present in \sref{periodic} and sums the series to
  get the final closed form of $\tbar$,
\begin{eqnarray}
\tbar(t,r,\theta,z,r',\theta',z') &=& 
 -\frac{1}{2\pi\theta_1 rr'\sinh u}\sum_{n=-\infty}^\infty
e^{-|\lambda|u+i\lambda(\theta-\theta')}\nonumber \\
&=& -\frac{1}{2\pi\theta_1 rr'\sinh u}
  \frac{\sinh(\frac{2\pi}{\theta_1}u)}{\cosh(\frac{2\pi}{\theta_1}u)
  - \cos\frac{2\pi}{\theta_1}(\theta -\theta')}\,.
\label{tbarfinal}\end{eqnarray}
 (Actually, Smith calculated the Wightman function (a Green 
function for the wave equation) rather than $\tbar$.  It differs 
only by replacing $t^2$ by $-(x^0-x^{0\prime})^2$.)

  \subsection{Nonhomogeneous problem; $t$--$\theta$--$z$ 
expansion}
The second method
 was employed by 
Smith \cite{smith} in dimension~$3$. 
 In our case (dimension~$4$)
it starts by observing that $\tbar$ (extended evenly in~$t$)
 satisfies the nonhomogeneous 
 Green-function equation 
 \begin{equation}\label{eq:drv2}
\frac{\partial^{2}\tbar}{\partial t^{2}} + 
 \frac{\partial^{2}\tbar}{\partial r^{2}} + 
 \frac{1}{r} \frac{\partial \tbar}{\partial r} + 
 \frac{1}{r^{2}} \frac{\partial^{2}\tbar}{\partial \theta^{2}} + 
 \frac{\partial^{2}\tbar}{\partial z^{2}} =
  \frac{2}{r} \delta (t) \delta (r - r^{'}) \delta (\theta - \theta^{'})
  \delta(z-z').
\end{equation}
 in all of~$\bo R^4$.
 The proof of \eref{eq:drv2} is a standard application of Green's 
identity combined with an image construction in the $t$ direction
 \cite{prenum}.

  To solve \eref{eq:drv2} we again use  a Fourier decomposition
  in $\theta$,
\begin{eqnarray}
\tbar(t,r,\theta,z) &=& \sum^{\infty}_{n=-\infty}
  e^{in\theta(\frac{2\pi}{\theta_{1}})} T_{n}(t,r,z), \\
T_{n}(t,r,z) &=& 
 \frac{1}{\theta_{1}} \int^{\theta_{1}}_{0}
 e^{-in\theta(\frac{2\pi}{\theta_{1}})} \tbar(t,r,\theta,z).
\end{eqnarray} 
 Again let $\lambda = \frac{2n\pi}{\theta_{1}}$, 
 and without loss of  generality take $\theta^{'} = 0$ and $z'=0$. 
 By substituting into (\ref{eq:drv2}), we get
\begin{equation}
\label{eq:t}
\frac{\partial^{2}T_{n}}{\partial t^{2}} + 
 \frac{\partial^{2}T_{n}}{\partial r^{2}} +
  \frac{1}{r} \frac{\partial T_{n}}{\partial r}
  - \frac{\lambda^{2}}{r^{2}}T_{n} + 
 \frac{\partial^2 T_{n}}{\partial z^2 }= 
 \frac{2}{\theta_1 r} \delta (t) \delta(r-r') \delta(z).
\end{equation} 
 The delta functions in $z$ and $t$ can be represented as
\begin{eqnarray}
\delta(z-z') &=& \frac{2}{\pi}\int^\infty_0 dk \cos kz \cos kz',\\ 
\delta(t-t') &=& \frac{2}{\pi}\int^\infty_0 d\omega' \cos \omega' t 
 \cos \omega' t',
\end{eqnarray} so we get
\begin{eqnarray}
\tilde{T}_n(\omega',r,k) &=& \int_0^\infty dt' \cos(\omega't')
 \int_0^\infty dz'\cos(kz') T_n(t',r,z'), \\                
T_n(t,r,z) &=& \frac{2}{\pi}\int^\infty_0 d\omega'
  \cos(\omega' t)\frac{2}{\pi}\int^\infty_0 dk \cos (kz)
  \tilde{T}_n (\omega',r,k).
\end{eqnarray} 
 Consequently, (\ref{eq:t}) becomes
\begin{equation}
-(\omega ^{'2}+k^2)\tilde{T}_n + \frac{\partial^{2}\tilde{T}_n}
 {\partial r^{2}} + \frac{1}{r} \frac{\partial \tilde{T}_n}
 {\partial r} - \frac{\lambda^{2}}{r^{2}}\tilde{T}_n =
  \frac{2}{\theta_1 r}\delta (r-r').
\label{eq:ttilde}\end{equation} 

  Let $\omega ^2 = \omega^{\prime2} + k^2$
 (not the same relation as in~\eref{eq:ZT}!). 
 The solution of \eref{eq:ttilde} has the form
\begin{equation}
\tilde{T}_n = 
 \cases
{C_{\omega}I_{|\lambda|}(\omega r) & for $ r < r' $, \cr
 D_{\omega}K_{|\lambda|}(\omega r) & for $ r > r'$.\cr}
 \end{equation} 
 From the continuity and jump conditions
 implementing the delta function in \eref{eq:ttilde}, we get that
\be
C_{\omega} = -\frac{2}{\theta_1} K_{|\lambda|}(\omega r'), \qquad
D_{\omega} = -\frac{2}{\theta_1} I_{|\lambda|}(\omega r').
\ee
  Therefore,
\begin{equation}
\tilde{T}_n (\omega ,r) = 
 -\frac{2}{\theta_1} I_{|\lambda|}(\omega r_{<})K_{|\lambda|}
 (\omega r_{>}),
\end{equation} and we get
\begin{eqnarray}
\tbar(t,r,\theta,z) &=& -\frac{2}{\pi^2 \theta_1}
 \sum^{\infty}_{n=-\infty} e^{in\theta(\frac{2\pi}{\theta_{1}})} 
 \nonumber\\
 &&\times \int^\infty_0 dk \int^\infty_0 d\omega' \cos(\omega' t)
  \cos (kz) I_{|\lambda|}(\omega r_{<})K_{|\lambda|}(\omega r_{>}).
\end{eqnarray} 

  Since $2\omega'\, d\omega' = 2\omega \,d\omega$
 and $\omega'=0$ when $\omega=k$,
 the formula can be rewritten
\begin{eqnarray}
\tbar(t,r,\theta,z) &=& -\frac{2}{\pi^2 \theta_1}
 \sum^{\infty}_{n=-\infty} e^{in\theta(\frac{2\pi}{\theta_{1}})}
  \int^\infty_0 \omega d\omega I_{|\lambda|}(\omega r_{<})
 K_{|\lambda|}(\omega r_{>}) \nonumber\\
&& \times \int^\omega_0 dk {\cos((\omega^2-k^2)^{1 \over 2} t)
  \cos (kz) \over (\omega^2-k^2)^{\frac{1}{2}}}\,.
\end{eqnarray}
  The $k$ integral yields $\frac{\pi}{2}J_0(\omega \zeta)$,
  where $\zeta \equiv \sqrt{z^2+t^2}$
 \cite[(3.876.7)]{GR}. Thus  by \cite[(6.578.11)]{GR}
\begin{eqnarray}
\tbar(t,r,\theta,z) &=& -\frac{1}{\pi\theta_1}
  \sum^{\infty}_{n=-\infty} e^{in\theta(\frac{2\pi}{\theta_{1}})} 
 \int^\infty_0 \omega d\omega I_{|\lambda|}(\omega r_{<})
 K_{|\lambda|}(\omega r_{>}) J_0(\omega\zeta) \nonumber\\
&=& -\frac{1}{\pi\theta_1} \sum^{\infty}_{n=-\infty} 
 e^{in\theta(\frac{2\pi}{\theta_{1}})} \times 
 \frac{e^{-i\pi /2}}{\sqrt{2\pi}} 
 {Q^{\frac{1}{2}}_{|\lambda|-\frac{1}{2}}(\cosh u) \over 
 rr'(\sinh u)^{\frac{1}{2}}},
\end{eqnarray}
  where
  \[\cosh u \equiv {r^2+r^{'2}+z^2+t^2 \over 2rr'}\] 
 and \cite[(8.754.4)]{GR}
 \[Q^{\frac{1}{2}}_{|\lambda|-\frac{1}{2}} (\cosh u) = 
 i\sqrt{\frac{\pi}{2\sinh u}} e^{-|\lambda|u}.\]
   Therefore,
\[
\tbar(t,r,\theta,z) = -\frac{1}{2\pi\theta_1 rr'\sinh u}
 \sum^{\infty}_{n=-\infty} e^{i\lambda\theta -|\lambda|u},
\]
 which agrees with the previous result~\eref{tbarfinal}.

{\sl Remark:\/}  Several other variations are possible (see 
\cite{ehk} for the 3-dimensional analogs).  
  One can 
solve \eref{eq:drv2} by a complete Fourier analysis in all 
variables; then whether the first or second method above develops 
depends on whether the first integral one evaluates is over the  
 variable $\omega'$ conjugate to~$t$ or the variable $\omega$
  conjugate to~$r$.
 Yet another approach \cite{DeuC,HK} 
 is to introduce an additional ``proper time'' 
coordinate and find a Fourier solution for the corresponding heat 
kernel or propagator; then integrating over the proper time segues 
back to the approach just described, but evaluating the $\omega'$ 
integral first yields a different representation.

 \subsection{Three-dimensional problem; positive mass}
Both methods described above can be used in three dimensions, 
 yielding the equivalent results \cite{smith}
\begin{eqnarray}\fl
T(t,r,r',\theta)&=& \frac{1}{\theta_1}
  \sum_{n=-\infty}^\infty e^{i\lambda\theta} 
 \frac{1}{\sqrt{rr'}}Q_{|\lambda|-\frac{1}{2}}(\cosh u_0)\label{3d1} \\
   &=& -\,\frac{1}{\pi\theta_1\sqrt{2rr'}}
 \int^\infty_{u_0} du \frac{1}{\sqrt{\cosh u -\cosh u_0}} \,
 \frac{\sinh (\frac{2\pi}{\theta_1}u)}{\cosh (\frac{2\pi}{\theta_1}u) 
-\cos(\frac{2\pi\theta}{\theta_1})}\,.
\label{3d2}\end{eqnarray} 
Here 
 \be \cosh u_0 = \frac{r^2 + r^{\prime2}+t^2}{2rr'}\,, \qquad
\lambda = \frac{2n\pi}{\theta_{1}}\,,
 \ee
 and $Q$ is a second-kind Legendre function.
 For details see \cite{smith,mai1}.

Formulas \eref{3d1} and \eref{3d2} cannot be simplified into an 
elementary closed form like \eref{tbarfinal}
 (just as a circular Bessel function is not elementary, but a 
spherical Bessel function is).
 However, \eref{3d2} clearly is related to \eref{tbarfinal};
 indeed, the former is a certain integral transform of the latter.
 To understand fully the relations between cylinder kernels in 
different dimensions, it is necessary to enlarge the discussion to 
allow the scalar field to have a mass \cite{GL}.
 That is, a term $m^2T$ is added to the right side of~\eref{TPDE}.
In the case of two spatial dimensions, this would lead via 
\eref{2Lap} to the equation 
 \be
\left( {\partial^2\over \partial{t^2}} + {\partial{^2}\over\partial{r^2}}
  +\frac1r \,{\partial{}\over\partial{r}} +\frac1{r^2}\, 
{\partial{}\over\partial{\theta^2}} - m^2\right)T =0
\label{masspde}  \ee
(or the corresponding equation for $\tbar$, or a similar 
nonhomogeneous equation like~\eref{eq:drv2}).
 But if we merely rename $m$ as $k$,  \eref{eq:drv2}
 is recognized as the result of Fourier-transforming in~$z$
 the four-dimensional massless field equation with spatial 
operator~\eref{3Lap}.
 Because we are working in Euclidean signature when we study 
cylinder kernels, a massive Klein--Gordon equation is the same 
thing as a Helmholtz equation in one higher dimension.

 It follows that by taking a Fourier transform of a massive 
cylinder kernel with respect to the mass, one obtains a cylinder 
kernel for the massless theory in one higher dimension.
 (The massive higher-dimensional case can be reached by using only 
a piece of the lower-dimensional mass in the transform 
 \cite[(2.5)]{GL}.)
 Conversely, by Fourier-transforming the four-dimensional $\tbar$
\eref{tbarfinal} in~$z$, and then setting the Fourier variable 
equal to zero, one obtains the three-dimensional massless~$\tbar$;
 and \eref{3d2} can be interpreted in this light after a change of 
variable from $z$ to~$u$ (using \eref{coshudef}, for example).

\section{The vacuum in periodic spaces}\label{periodic}

In summary, we have (in total dimension 4)
 for the cone with angular periodicity $\theta_1$
\be
\tbar_{\theta_1}(t,r,r',\theta,z) = 
 -\,\frac1 {2\pi\theta_1 rr'\sinh u}\, 
 {\sinh\left(2\pi  u\over \theta_1\right) 
\over
 \cosh \left(2\pi  u\over \theta_1\right)  - 
 \cos \left(2\pi \theta \over \theta_1\right) 
}\,,
 \label{fullTbar}\ee
 where $u$ is defined by any of the equivalent formulas
 \be
 u= -\ln {r_2-r_1 \over r_2+r_1}\,, 
\label{udef} \ee
 \[
\qquad r_1= \sqrt{(r-r')^2 + z^2+ t^2}\,,\quad
r_2= \sqrt{(r+r')^2 + z^2+ t^2}\,; 
   \]
 \be 
 2rr'\cosh u = r^2 + r'^2 + z^2 + t^2;
 \label{coshudef}\ee
 \be
 2rr'\sinh u = \sqrt{[r^2+r'^2 + z^2 +t^2]^2 -4r^2r'^2}\,;
\label{sinhudef} \ee
 \be
 4rr'\sinh^2\left(\frac u2\right)= (r-r')^2 + z^2 +t^2.
\label{sinhuhalf} \ee
(Without loss of generality we have set $t'$, $z'$, and $\theta'$ 
equal to $0$.
 When derivatives with respect to primed coordinates are needed, 
one simply replaces $z$ by $z-z'$, etc., first.)

 When $\theta_1=2\pi$, (\ref{fullTbar}) reduces to the formula for 
Minkowski space in cylindrical coordinates,
 \begin{eqnarray}
 \tbar_{2\pi} &=& -\, \frac1{4\pi^2  rr'} \,\frac1{\cosh u - \cos \theta}
\label{fullflatTbar} \\
 &=&  -\frac{1}{2\pi^2 (r^2+r'^2+t^2+z^2-2rr'\cos(\theta))}\,.
 \label{standardTbar} \end{eqnarray}
 (The two forms are shown equivalent by  (\ref{coshudef}) and the law of 
cosines.)
 When $\theta_1\to\infty$, (\ref{fullTbar})  becomes the formula for 
the infinite-sheeted Dowker space-time,
 \be
 \tbar_\infty = -\,\frac1{2\pi^2 rr' \sinh u}\, {u\over u^2+ \theta^2}\,.
\label{fullDowTbar} \ee

 $\tbar_{\theta_1}$ can be recovered from $\tbar_\infty$ as a 
periodic image sum:
 after suppressing the irrelevant coordinates,
  \be \tbar_{\theta_1}(\theta-\theta') = 
 \sum_{n=-\infty}^\infty \tbar_\infty(\theta-\theta'+ n\theta_1).
 \label{anglesum}\ee
 In particular,
  \be \tbar_{2\pi}(\theta-\theta') = 
 \sum_{n=-\infty}^\infty \tbar_\infty(\theta-\theta'+2\pi n).
 \label{freeanglesum}\ee
 In terms of normal modes, $\tbar_{2\pi}$ is a sum over angular 
momentum quantum number, while $\tbar_\infty$ is an integral over 
angular momentum.
 $\tbar_\infty$ can be found by separation of variables 
 (integration over modes) in a slight variation of the calculation 
in the previous section.
   Formula \eref{anglesum} can be verified, or discovered, in 
  \emph{Mathematica} or by complex analysis; 
indeed, the entire theory of vacuum stress in cones and wedges 
could be developed from formula \eref{fullDowTbar} in this way.
  The Dowker Green function 
 $\tbar_\infty$ is physically the most elementary;
  it tells how to diffract a path off the isolated conical 
singularity.
 The additional effects of periodicity, and hence of reflecting 
wedge walls, can be obtained from $\tbar_\infty$  
 by images, or  summation over paths, appropriate to each geometry 
in question.
Of course,  these remarks extend to the energy density and the rest 
of the stress tensor, and also
  to other Green functions, not just the cylinder kernel.

 The sum (\ref{anglesum}) has a structure similar to 
the formula for 
 $\tbar$ in a  universe that is periodic in one direction with 
period~$L_1\,$:
  \be
  \tbar_{L_1}(x-x') = \sum_{n=-\infty}^\infty \tbar_0(x-x'+nL_1).
 \label{periodsum} \ee
Again,
  $\tbar_{L_1}$ is a Fourier sum, $\tbar_0$  a Fourier integral.
 The algebra leading from (\ref{fullDowTbar}) to (\ref{fullTbar}) 
according to (\ref{anglesum}) is very similar to that leading from 
 (\ref{T0}) to a closed form for (\ref{periodsum}).

There is, of course, a major glitch in the analogy of 
$\tbar_\infty$ and $\tbar_{\theta_1}$ to $\tbar_0$ and $\tbar_{L_1}\,$, 
respectively:  
it is $\tbar_{2\pi}$ (not $\tbar_\infty$) that, like $\tbar_0\,$,
  is the free-space Green function that gives 
the  zero-point energy density that must be
  subtracted from the energy of any other configuration. 
  A periodic space has nontrivial Casimir 
energy, inherent in $\tbar_{L_1}$ after the subtraction.
In the early stages of the universe, this vacuum energy should affect 
 the   cosmological expansion.

 Actually, this last conclusion is not universally accepted.
 The critique  runs somewhat as follows:
 A real physicist is interested in the scalar 
field only as a model of the electromagnetic field.
 The (experimentally verified) electromagnetic Casimir energy 
 is the energy of 
interaction of fluctuations of the electrons in the bounding conductors
 (the long-range limit of the van der Waals force).
  The field energy is just a bookkeeping device.
  But in a periodic universe
 (or a higher-dimensional closed universe)
  there are no 
boundaries, no fluctuating electrons.  There is no  
experimental evidence that cosmological vacuum energy exists.
 (The observed dark energy may have something to do with zero-point 
energy, despite the notorious factor $10^{120}$,
 but it has nothing to do
 (in the absence of small extra dimensions, at least~\cite{kimKK})
  with the renormalized Casimir energy, 
which has a rest frame and is unobservably small at the present 
epoch.)

Nevertheless, the mathematics (and indirectly confirmed physics) of 
quantum field theory unambiguously says that the vacuum energy density
 of a periodic universe differs from that of infinite space by a 
locally  finite amount in precise analogy to how the vacuum energy 
density of infinite space (looked at in polar coordinates) differs 
from that of Dowker space.
 In each case the increment is the difference between a Fourier
series and a Fourier integral over the same spectral integrand.
   If you believe in the stress tensor of quantum field
  theory, this picture  is totally consistent and unsurprising
 (except for the enduring mystery of what happens to the zero-point 
energy).                        
 But if you don't, you are forced into an untenable position:
\begin{itemize}           
 \item  To calculate vacuum energy in the periodic universe,
  you must ignore the 
(mathematically appropriate) Fourier sum in favor of the integral
 (since you don't believe vacuum energy can exist in the absence of 
van der Waals sources).
 \item  In polar coordinates, you must use the sum to get 
the right answer for empty Euclidean space.  The integral gives 
something else, the energy density of the Dowker manifold.
\end{itemize} 
 We see no possibility of a theoretical justification for this ad 
hoc switch of point of view.
 (Of course, this is not the same as an experimental or 
observational verification of the theory.)

 \section{Numerics}

We now present some vacuum energy calculations for various cone and 
wedge spaces, based on the cylinder kernel for a cone as given
  in~\eref{fullTbar} and alternatively in~\eref{anglesum},
and executed in {\sl Mathematica}.
Most of these results were previously presented 
in~\cite{tthesis}.

In this section we will work on diagonal, 
 with $r'=r$, $z=0$, and $\theta=0$, but with $t>0$. 
 This is the traditional exponential ultraviolet cutoff.
 For flat space, we have 
\be
T_{00}=\frac{3}{2\pi^2 t^4}, \quad
 T_{rr}=T_{\bot\bot}=T_{zz}=\frac{1}{2\pi ^2 t^4}\,,
\label{0ptstress}\ee
 and of course we expect to subtract this ``zero-point stress'' 
from 
the stress tensor in any other configuration to get a physically 
meaningful quantity;
 this has been done in all the plots below, and 
in~\eref{wedgeTbar}.
Because $t^{4} T_{\mu\nu}$ (with $z=0$) depends on $r$ and $t$ 
only in the ratio $r/t$,
the plot for $t=1$ actually serves for all positive values 
of~$t$, as the axis labels indicate.

 In the plots, solid curves are for the 
ultraviolet cutoff $t\neq 0$ and dashed curves are for $t=0$.
 When the cutoff is thus removed, the energy density and pressures 
are finite at each point in the interior of~$\Omega$ but (usually)
 develop nonintegrable singularities at the boundaries 
($\theta\to0$, $\theta\to\theta_0\,$, or $r\to0$).
 Note that when a cutoff is present ($t>0$), 
 the key quantity $u$ is strictly positive even when all the 
coordinates are on diagonal ($r=r'$, $z=z'=0$, $\theta=\theta'=0$);
 therefore, all the quantities being plotted remain finite at the 
wedge plates, although the plots may give the impression of vertical 
asymptotes there. 

 When multiple traces are shown in the same panel, blue, red, and 
(when present) yellow curves correspond to the cases in the order 
listed in the caption.  To facilitate viewing on a monochrome 
device, we indicate in each case whether the corresponding ordering 
is from top to bottom or bottom to top in the bulk of the figure.
Also, multiple plots within one figure are sometimes plotted with different
horizontal and vertical scales, in order to better show the behavior
of the functions. In many cases, the energy density has been replotted
in a separate graph to show the behavior for small $r$ more clearly.

 Plots of the energy density and pressure    in 
the Dowker space with $\beta=0$ ($\xi=\frac14$), obtained
  from~\eref{fullDowTbar}, are given in 
 figure~\ref{figure1}.
  Figure~\ref{figure1b} displays 
 the additional  terms in  the energy 
density and pressure   when $\beta\neq 0$.
 Because these correction terms are proportional to $\beta\equiv 
\xi-\frac14$, it suffices to plot the case $\beta=1$.
\begin{figure}[h]
\begin{center}$
\begin{array}{cc}
\includegraphics[width=8cm]{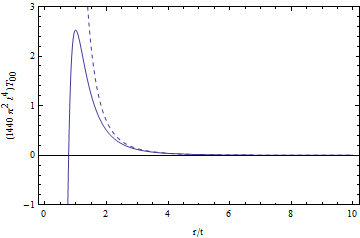} &
\includegraphics[width=8cm]{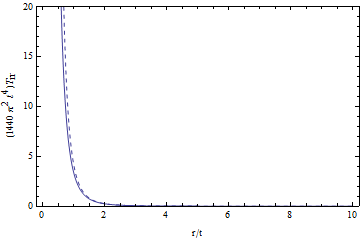}\\
$(a) Energy density$ & $(b) Radial pressure$\\
\includegraphics[width=8cm]{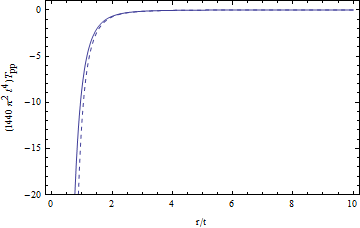} &
\includegraphics[width=8cm]{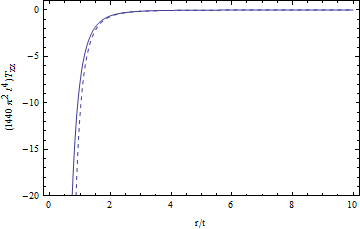}\\
$(c) Tangential pressure$ & $(d) Axial pressure$\\
\end{array}$
\end{center}
\caption{Plot of the Dowker energy density and pressure as a 
 function of $r$ with $\xi=\frac{1}{4}$.}
\label{figure1}
\end{figure}
\begin{figure}[h]
\begin{center}$
\begin{array}{cc}
\includegraphics[width=8cm]{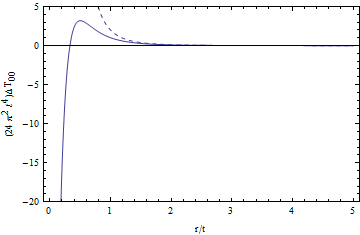} &
\includegraphics[width=8cm]{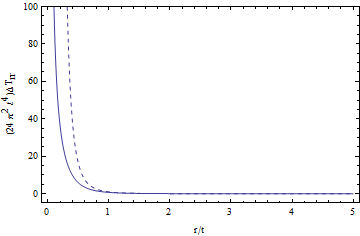}\\
$(a) Energy density correction$ & $(b) Radial pressure correction$\\
\includegraphics[width=8cm]{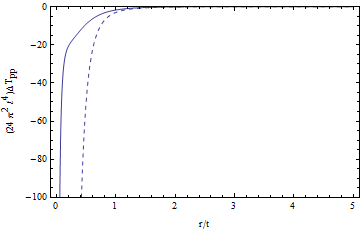} &
\includegraphics[width=8cm]{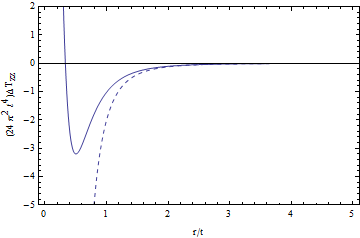}\\
$(c) Tangential pressure correction$ & $(d) Axial pressure correction$\\
\end{array}$
\end{center}
\caption{Plot of the Dowker energy density and pressure 
curvature-coupling corrections as a 
 function of $r$ with $\beta=1$.}
\label{figure1b}
\end{figure}

 In figures \ref{figure2mis} and \ref{figure2ext}
 the energy density and pressure  are presented 
for  cone spaces with various cone angles.
 Notice that the overall sign of the energy density function changes at 
$\theta_1=2\pi$.   Plots for $\theta_1=10 000\pi$ 
match very well with plots for  Dowker space, as
expected since  Dowker space is the large-period limit of the cones. 
The behavior of the energy density near $r=0$ is illustrated in figure \ref{figure2b}.
\begin{figure}[h]
\begin{center}$
\begin{array}{cc}
\includegraphics[width=8cm]{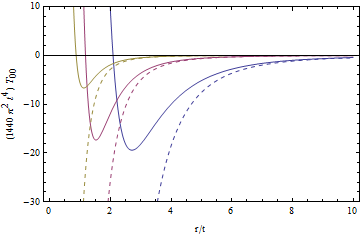} &
\includegraphics[width=8cm]{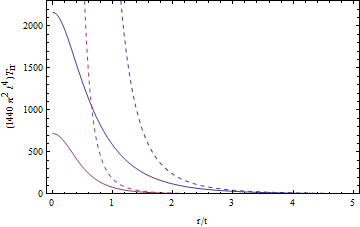}\\
$(a) Energy density$ & $(b) Radial pressure$\\
\includegraphics[width=8cm]{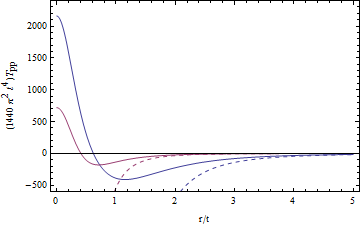} &
\includegraphics[width=8cm]{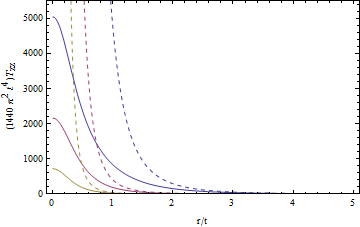}\\
$(c) Tangential pressure$ & $(d) Axial pressure$\\
\end{array}$
\end{center}
\caption{Energy density and pressure for $\xi= 1/4$ and cone angles 
$\theta_1=\pi/4$, $\pi/2$, and $\pi$ (respectively blue, red, 
yellow, bottom to top in (a) and (c), and top to bottom in (b) and (d)).
(Some plots for $\theta_1=\pi$ are missing because of a 
numerical problem.)}
\label{figure2mis}
\end{figure}
\begin{figure}[h]
\begin{center}$
\begin{array}{cc}
\includegraphics[width=8cm]{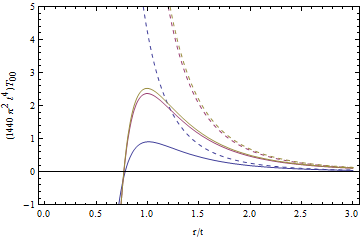} &
\includegraphics[width=8cm]{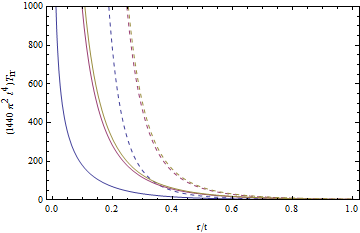}\\
$(a) Energy density$ & $(b) Radial pressure$\\
\includegraphics[width=8cm]{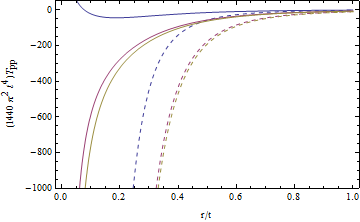} &
\includegraphics[width=8cm]{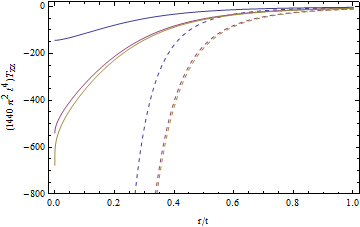}\\
$(c) Tangential pressure$ & $(d) Axial pressure$\\
\end{array}$
\end{center}
\caption{Energy density and pressure for $\xi= 1/4$ and cone angles 
$\theta_1=2.5\pi$, $8\pi$, and $10 000\pi$ (respectively blue, 
red, 
yellow, bottom to top in (a) and (b), and top to bottom in (c) and (d)).}
\label{figure2ext}
\end{figure}
\begin{figure}[h]
\begin{center}$
\begin{array}{cc}
\includegraphics[width=8cm]{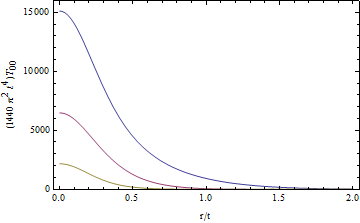} &
\includegraphics[width=8cm]{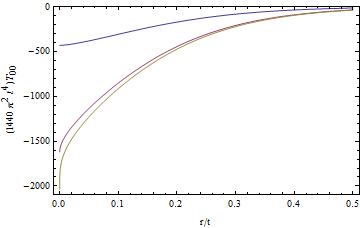}\\
$(a) Energy density$ & $(b) Energy density$
\end{array}$
\end{center}
\caption{Energy density for $\xi= 1/4$   and cone angles 
 (a) $\theta_1=\pi/4$, $\pi/2$, and $\pi$ (respectively blue, 
red, 
yellow, top to bottom), and 
 (b) $\theta_1=2.5\pi$, $8\pi$, and $10 000\pi$ (top  to bottom). 
Behavior near $r=0$.}
\label{figure2b}
\end{figure}

 The curvature-coupling correction terms to the energy density 
and pressure
 are presented in figures \ref{figure3} and~\ref{figure3b}  for 
various cone angles.
 The energy density for the conformal case, where $\xi=1/6$, 
 is plotted in 
 figure \ref{figure4} and compared with the case $\xi=1/4$.
\begin{figure}[h]
\begin{center}$
\begin{array}{cc}
\includegraphics[width=8cm]{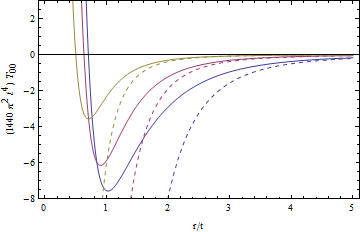} &
\includegraphics[width=8cm]{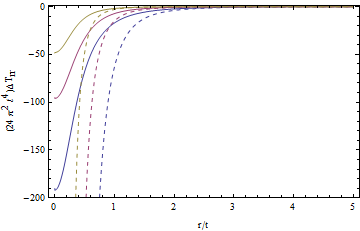}\\
$(a) Energy density correction$ & $(b) Radial pressure correction$\\
\includegraphics[width=8cm]{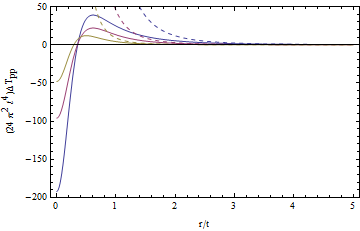} &
\includegraphics[width=8cm]{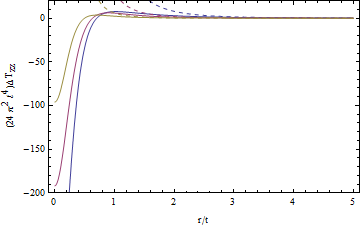}\\
$(c) Tangential pressure correction$ & $(d) Axial pressure correction$\\
\end{array}$
\end{center}
\caption{Energy density and pressure correction terms for $\beta=1$ 
 and cone angles 
 $\theta_1=\pi/4$,  $\pi/2$, and $\pi$ (bottom to top).}
\label{figure3}
\end{figure}
\begin{figure}[h]
\begin{center}
\includegraphics[width=8cm]{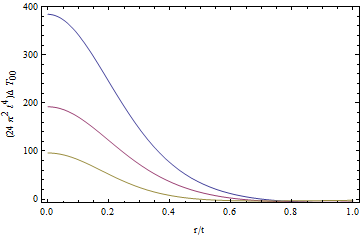}
\end{center}
\caption{Energy density correction term for $\beta=1$ 
 and cone angles 
 $\theta_1=\pi/4$,  $\pi/2$, and $\pi$ (top to bottom). Behavior 
near $r=0$.}
\label{figure3b}
\end{figure}
 \begin{figure}[h]
\begin{center}$
\begin{array}{cc}
\includegraphics[width=8cm]{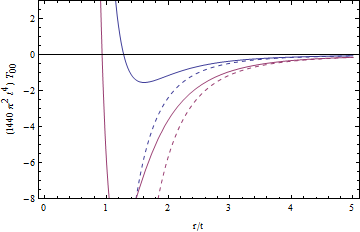} &
\includegraphics[width=8cm]{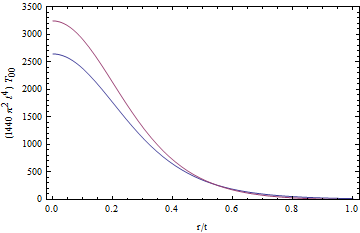}\\
$(a) Energy density$ & $(b) Energy density$\\
\end{array}$
\end{center}
\caption{Energy density for $\xi=1/6$ and $\xi=1/4$ (top to bottom in (a),
 and bottom to top in (b)), with cone angle $\theta_1=0.8\pi$. Behavior near
$r=0$ is shown in (b).}
\label{figure4}
\end{figure}

In figures \ref{coneang1} and~\ref{coneang2}
we  present the energy density and pressure for the cone space,
 and the correction terms, 
 as functions of the cone size $\theta_1$ with fixed $r=1$.
\begin{figure}[h]
\begin{center}$
\begin{array}{cc}
\includegraphics[width=8cm]{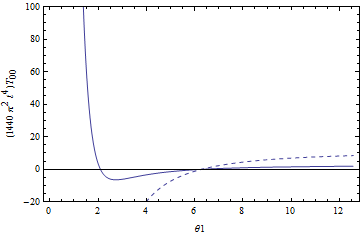} &
\includegraphics[width=8cm]{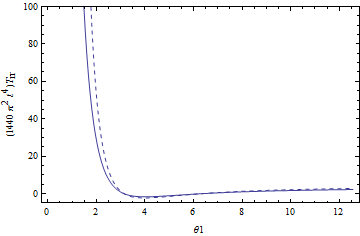}\\
$(a) Energy density$ & $(b) Radial pressure$\\
\includegraphics[width=8cm]{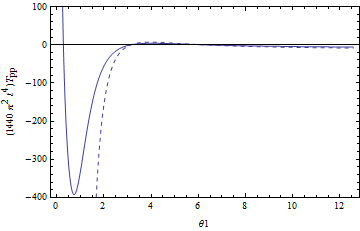} &
\includegraphics[width=8cm]{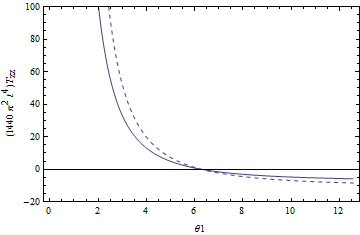}\\
$(c) Tangential pressure$ & $(d) Axial pressure$\\
\end{array}$
\end{center}
\caption{Energy density and pressure for the cone space as a function of $\theta_1$,
with $\beta=0$ and $r=1$.}
\label{coneang1}
\end{figure}
\begin{figure}[h]
\begin{center}$
\begin{array}{cc}
\includegraphics[width=8cm]{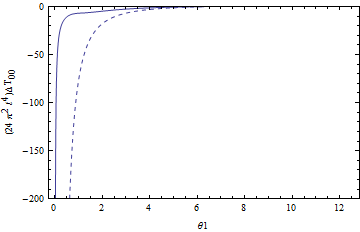} &
\includegraphics[width=8cm]{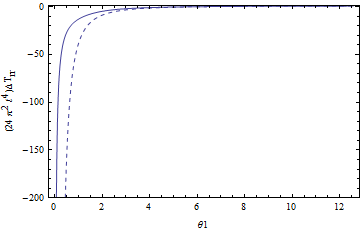}\\
$(a) Energy density correction$ & $(b) Radial pressure correction$\\
\includegraphics[width=8cm]{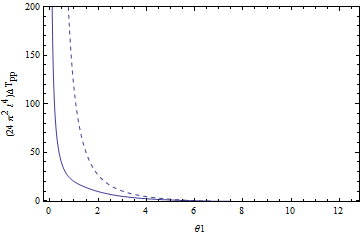} &
\includegraphics[width=8cm]{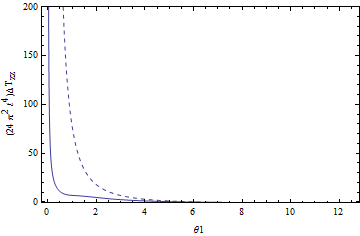}\\
$(c) Tangential pressure correction$ & $(d) Axial pressure correction$\\
\end{array}$
\end{center}
\caption{Energy density and pressure curvature-coupling 
corrections for 
the cone space as a function of $\theta_1\,$,
with $r=1$.}
\label{coneang2}
\end{figure}

For a Dirichlet wedge with opening angle $\theta_0\,$, we have an 
 infinite sum over positive and negative images in Dowker space, 
 and the resulting expression for $\tbar$ is
\begin{eqnarray}
-\,\frac{1}{4\pi\theta_0 rr' \sinh(u)}&
 \left(\frac{\sinh(\pi u/\theta_0)}
 {\cosh(\pi u/\theta_0)-\cos(\pi(\theta-\theta')/\theta_0)}
\right. \nonumber\\ 
 &\left. -\,\frac{\sinh(\pi u/\theta_0)}{\cosh(\pi u/\theta_0)
 -\cos(\pi(\theta+\theta')/\theta_0)}\right) \nonumber\\
&{} +\frac1{4\pi^2  rr'} \,\frac1{\cosh u - \cos \theta}\,.
 \label{wedgeTbar}\end{eqnarray}
 The energy density now becomes a function of $\theta$ as well 
as~$r$.
 Figures \ref{figure5} and \ref{figure6} show the energy density 
 as a function of $\theta$ for various values of $\xi$, 
$\theta_0\,$,  and $r$.
Figures \ref{figure5}(b) and \ref{figure6}(b) show that when the 
cutoff is removed, the conformal energy density is independent of 
the angle (a fortiori, not divergent at the wedge plates),
  but at finite cutoff  the function deviates strongly from this 
  limiting behavior near the boundary.
 Both panels of figure \ref{figure5} show that the cutoff function 
is very far from the limit function when $r$ is rather small.
Figures \ref{figure5b} and \ref{figure6b} show  that the energy density
remains finite at the boundaries.
Figures \ref{figure7} and~\ref{figure7b} show the energy density 
as a function of $r$
 for various values of $\xi$, $\theta_0\,$, and $\theta$. 
\begin{figure}[h]
\begin{center}$
\begin{array}{cc}
\includegraphics[width=8cm]{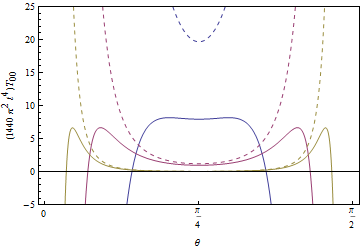} &
\includegraphics[width=8cm]{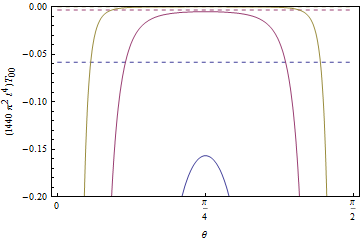}\\
$(a) Energy density$ & $(b) Energy density$
\end{array}$
\end{center}
\caption{Energy density in a wedge as a function of $\theta$ with 
 $\theta_0=\pi/2$ and (a) $\xi=1/4$, for $r$ at $2$, $4$, and $8$
 (top to bottom); 
 (b) $\xi=1/6$, for $r$ at $4$, $8$, and $16$ (bottom to top).}
\label{figure5}
\end{figure}
\begin{figure}[h]
\begin{center}$
\begin{array}{cc}
\includegraphics[width=8cm]{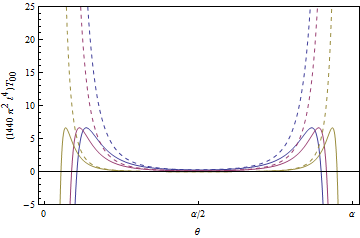} &
\includegraphics[width=8cm]{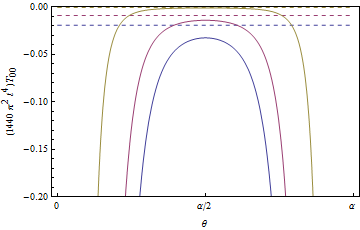}\\
$(a) Energy density$ & $(b) Energy density$
\end{array}$
\end{center}
\caption{Wedge energy density as a function of $\theta$ with 
$r=8$ and 
 $\theta_0$ values $\pi/3$, $2\pi/5$, and $2\pi/3$ for (a) $\xi=1/4$
 (top to bottom); 
 (b) $\xi=1/6$ (bottom to top).}
\label{figure6}
\end{figure}
\begin{figure}[h]
\begin{center}$
\begin{array}{cc}
\includegraphics[width=8cm]{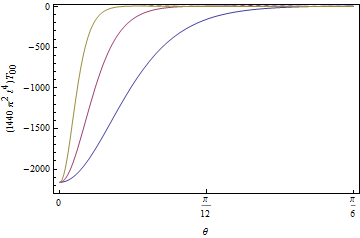} &
\includegraphics[width=8cm]{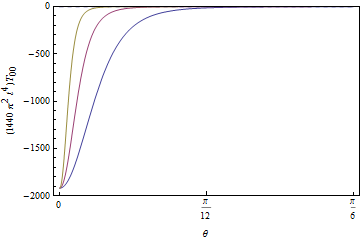}\\
$(a) Energy density$ & $(b) Energy density$
\end{array}$
\end{center}
\caption{Wedge energy density as a function of $\theta$ with 
 $\theta_0=\pi/2$ and (a) $\xi=1/4$, for $r$ at $2$, $4$, and $8$
 (bottom to top); 
 (b) $\xi=1/6$, for $r$ at $4$, $8$, and $16$ (bottom to top). 
Behavior near $\theta=0$.}
\label{figure5b}
\end{figure}
\begin{figure}[h]
\begin{center}$
\begin{array}{cc}
\includegraphics[width=8cm]{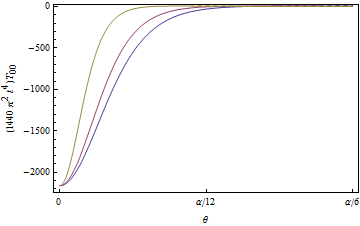} &
\includegraphics[width=8cm]{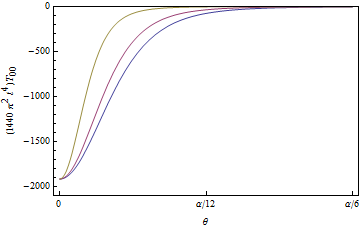}\\
$(a) Energy density$ & $(b) Energy density$
\end{array}$
\end{center}
\caption{Wedge energy density as a function of $\theta$ with 
$r=8$ and 
 $\theta_0$ values $\pi/3$, $2\pi/5$, and $2\pi/3$ for (a) $\xi=1/4$
 (bottom to top); 
 (b) $\xi=1/6$ (bottom to top). Behavior near $\theta=0$.}
\label{figure6b}
\end{figure}
 \begin{figure}[h]
\begin{center}$
\begin{array}{cc}
\includegraphics[width=8cm]{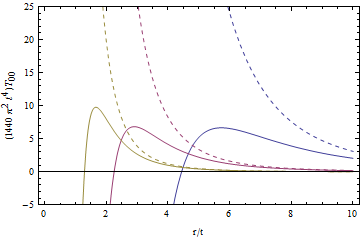} &
\includegraphics[width=8cm]{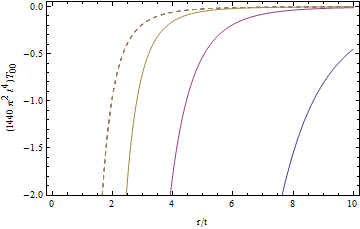}\\
$(a) Energy density$ & $(b) Energy density$
\end{array}$
\end{center}
\caption{Wedge energy density as a function of $r$ with 
$\theta_0=\pi/2$ and 
 $\theta$ values $\pi/16$, $\pi/8$, and $\pi/4$ for (a) $\xi=1/4$
 (top to bottom); 
 (b) $\xi=1/6$ (bottom to top).}
\label{figure7}
\end{figure}
 \begin{figure}[h]
\begin{center}$
\begin{array}{cc}
\includegraphics[width=8cm]{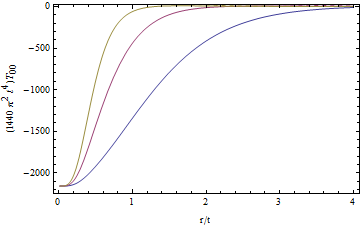} &
\includegraphics[width=8cm]{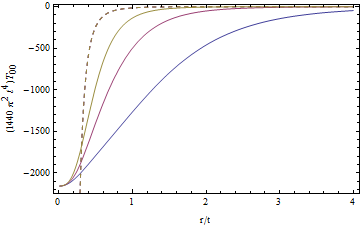}\\
$(a) Energy density$ & $(b) Energy density$
\end{array}$
\end{center}
\caption{Wedge energy density as a function of $r$ with 
$\theta_0=\pi/2$ and 
 $\theta$ values $\pi/16$, $\pi/8$, and $\pi/4$ for (a) $\xi=1/4$
 (bottom to top); 
 (b) $\xi=1/6$ (bottom to top). Behavior near $r=0$.}
\label{figure7b}
\end{figure}
 
Numerically, there is no qualitative difference between ``good'' 
and ``bad'' wedge angles, despite the great difference between 
them with regard to solvability by the classic method of images.
The exception is the case $\theta_0=\pi$ (for which the ``cone'' 
is flat space),  where the ``wedge'' energy is just that of a 
single reflecting plane, expressed in polar coordinates.

 \section{Conclusions and outlook}

 There is a dual relationship between infinite (Euclidean) space
 and finite flat space (a circle or torus):
 Green functions for   the infinite case can be obtained from the 
finite case by taking a limit, in which a Fourier series becomes a 
Fourier transform, or the finite, periodic case can be obtained 
from the infinite one as an infinite sum over image sources.
 A similar relationship exists in a two-dimensional space that is 
locally flat except at a single conical singularity (hence in a 
higher-dimensional space with a singular axis), except that the 
Fourier analysis is now with respect to the angular coordinate. 
The role of the infinite space is now played by an infinite-sheeted 
singular surface, whose importance has been particularly stressed 
by Dowker.  Among the continuum of periodic factor spaces (cones) 
of the Dowker manifold with respect to periodic image sums, one is 
our familiar, nonsingular Euclidean space. From the cone of 
arbitrary angle a further image sum, with respect to a single 
reflection, yields the Green functions for a perfectly reflecting 
wedge  in Euclidean space, with an arbitrary opening angle.

 As shown by Lukosz and Smith, the particular Green function that 
is most useful for calculating the vacuum expectation value of the
 stress tensor in scalar quantum field theory can be found in 
closed form by rather elementary methods in all these spaces ---
 Dowker space, cones, and wedges.  The vacuum energy density (and 
pressure) in such systems consists of three components:  the effect 
of the central singularity present already in the Dowker space,
 the effect of periodicity associated with a cone, and the specific 
effect of the reflecting edges of the wedge.   The first two of 
these are independent of the angular coordinate inside the wedge.
From this point of view, Euclidean space is itself a cone, whose 
periodicity energy precisely cancels the energy of the (erstwhile) 
 central singularity.  This periodicity energy is closely analogous 
to that of a torus relative to infinite space, which has sometimes 
been questioned because of the absence of sources.

We have presented numerous plots of the (renormalized) energy 
density and pressures in these systems, both with and without an 
exponential ultraviolet cutoff.  
 In some circumstances a cutoff can be regarded as a mathematical 
 device that can be completely removed at the end.  However,
 alternatively a finite cutoff can be viewed as a small step toward 
 a realistic physical model of the boundary material.
 In that case the standards for physical acceptability of the 
renormalized cutoff stress tensor are higher.
Recent work on plane boundaries \cite{qfe09,qfe11} 
 indicates that the exponential cutoff introduces unphysical behavior
 in the energy density (but not the pressure) very near the boundary,
 and suggests that a better model of both energy and pressure could be
 obtained by replacing that cutoff (which corresponds to a 
separation of the arguments of the Green function in the imaginary 
 time direction) 
 by a point separation in a ``neutral'' direction.
 In the present cylindrical situation that would mean a separation 
in the axial direction, and that prescription is currently under 
investigation, starting in~\cite{tthesis}.

  \ack
We are grateful to the editors of this Dowkerfest volume for the 
invitation to write this article, and to the organizers of the 
Miltonfest conference in Norman, Oklahoma, in 2010 for the 
opportunity to present a preliminary account there.
 As a corollary, we are ever grateful to Stuart Dowker and Kim 
Milton for their inspiration and counsel.
 We thank R. Jackiw and F. D. Mera for comments.
 This work was supported by NSF Grants PHY-0554849 and PHY-0968269
 and by a one-semester postdoctoral appointment for J. Wagner in 
the Texas A\&M Mathematics Department.

 \end{document}